\documentstyle[twoside,fleqn,espcrc2]{article}

\begin{document}
\hyphenation{Super-Kamiokande}
\input psfig

\title{ASTROPHYSICAL NEUTRINOS: 20th CENTURY AND BEYOND}

\author{J. N. BAHCALL, IUPAP Centennial Lecturer\address{Institute for
        Advanced Study, Princeton, NJ 08540, USA\\E-mail:
        www.sns.ias.edu/$\sim$jnb}}

\begin{abstract}
I summarize the first four decades of solar neutrino research and
suggest what may be possible to learn with extragalactic neutrinos and with
solar neutrinos in the next decade.
\end{abstract}

\maketitle

\section{Introduction}
\label{sec:introduction}

I was asked by Art McDonald to give one of the opening
talks on the subject `` Neutrino Astrophysics in the 20th Century and
Beyond.''  Feeling very honored, I readily accepted. But, as I started
thinking about what I should say to so many knowledgeable friends, I
got really worried.  I realized that it would be impossible in
twenty-five minutes to discuss intelligently all of neutrino
astrophysics of the 20th century. There is just too much important
material to cover even if I spoke very fast, unintelligibly fast, and
even if I did not say anything about our goals for the future.

So, I decided to limit my remarks to two specific topics: solar
neutrinos and extragalactic neutrinos. I will not say anything about
the enormous achievements in the prediction and detection of supernova
neutrinos and in the calculations of neutrino cooling processes for
stars that are not exploding.  I will also not discuss the role of
neutrinos in Big Bang nucleosynthesis nor in cosmology. There are
lots of grand things to say about these subjects, and many other
topics in neutrino astrophysics, but I will not discuss them today.

I will take a somewhat historical approach and emphasize
those aspects of the development of our subject which may help guide
our thinking about what we should do in the future. I will begin with
solar neutrinos and then switch abruptly to extragalactic neutrinos.

\pagebreak

\begin{figure}[!htb]
\centerline{\psfig{figure=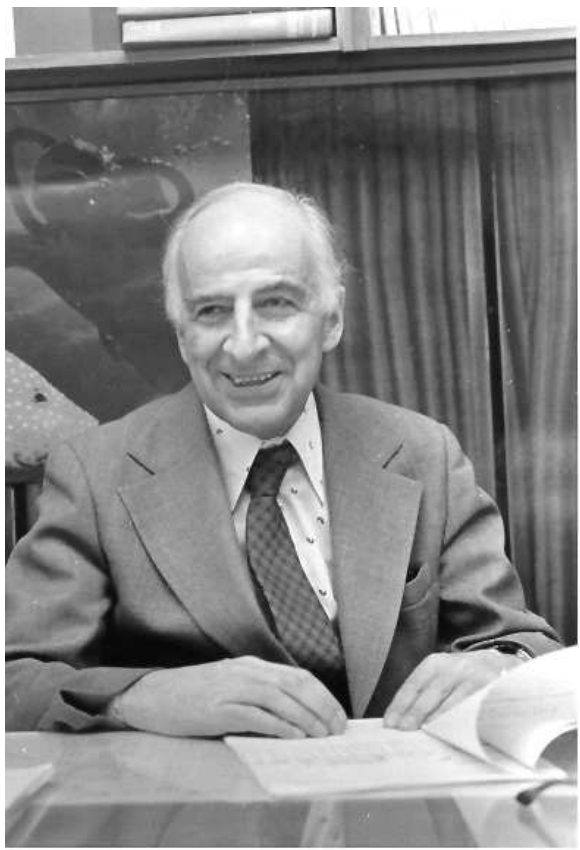,width=3in}}
\vglue-.3in
\caption[]{Bruno Pontecorvo wrote in 1967: `From the point of view of
detection possibilities, an ideal object is the sun.'
Figure courtesy of S. Bilenky .\label{fig:pontecorvo}}
\end{figure}

\section{Solar neutrinos}
\label{sec:solarneutrinos}

\subsection{Bruno Pontecorvo and Ray Davis}
\label{sec:brunoandray}

I want to begin by paying tribute to two of the great scientists and pioneers
of neutrino astrophysics, Ray Davis and Bruno Pontecorvo. Bruno first
suggested using chlorine as a detector of neutrinos in a Chalk River
report written in 1946. Ray followed Bruno's suggestion and the
careful unpublished feasibility study of Louie Alvarez. Using with
 care and skill a chlorine detector and reactor neutrinos, Ray
showed in 1955-1958 that $\nu_e$ and $\bar\nu_e$ were different. About
a decade later, Ray first detected solar neutrinos, laying the
foundation for the studies we shall hear about today.

In 1967, one year before the first results of Ray's chlorine solar
neutrino experiment were announced, Bruno published a prophetic paper
entitled: `Neutrino Experiments and the Problem of Conservation of
Leptonic Charge' [Zh. Exp. Teor. Fiz. 53, 1717 (1967)]. In this paper,
Bruno suggested many different experiments that could test whether
leptonic charge was conserved. The grandchildren of most of these
experiments are being discussed in this conference, Neutrino 2000.

Bruno included a short section in his paper that he called
`Oscillations and Astronomy.' In this section, Bruno wrote: ``From the
point of view of detection possibilities, an ideal object is the
sun,'' What a wonderfully contemporary statement!

Bruno, like most particle physicists of the 1960's and perhaps 1970's
  and 1980's, did not believe astrophysical calculations could be
  reliable. He wrote in this same section on oscillations and
  astronomy: ``Unfortunately, the weight of the various thermonuclear
  reactions in the sun, and the central temperature of the sun are
  insufficiently well known in order to allow a useful comparison of
  expected and observed solar neutrinos, from the point of view of
  this article.'' [This was 30 years before the precise confirmation
  of the standard solar model by helioseismology.] To support his
  claim, Bruno referenced only his 1946 Chalk River report, which
  mentioned the sun in just two sentences.  Bruno did cite our
  calculations of the solar neutrino fluxes elsewhere in his 1967
  paper, but they seem not to have affected his thinking.

\begin{figure}[!htb]
\hglue-.27in\psfig{figure=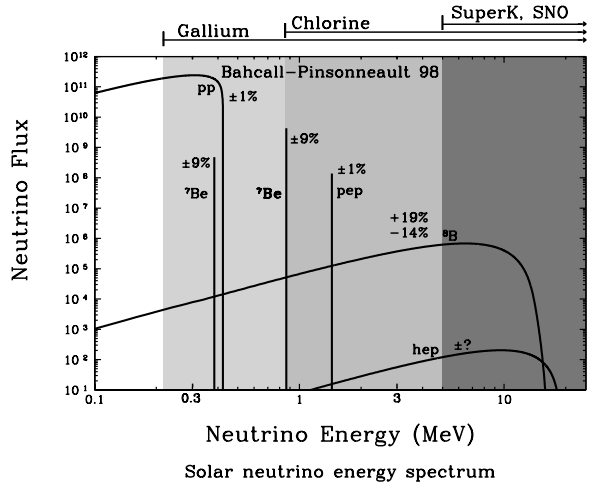,width=3.6in,angle=270}
\vglue-.3in
\caption[]{The energy Spectrum of neutrinos from the pp chain of
interactions in the Sun, as predicted by the standard solar model.
Neutrino fluxes from continuum sources (such as $p-p$ and $^8$B) are
given in the units of counts per ${\rm cm^2}$ per second.  The
percentage errors are the calculated $1\sigma$ uncertainties in the
predicted fluxes.  The $p-p$ chain is responsible for more than 98\% of
the energy generation in the standard solar model. Neutrinos produced
in the carbon-nitrogen-oxygen CNO chain are not important
energetically and are difficult to detect experimentally. The arrows
at the top of the figure indicate the energy thresholds for the
ongoing neutrino experiments.  This spectrum is from BP98: J. N. Bahcall,
S. Basu, and M. H. Pinsonneault, Phys. Lett. B, 433, 1 (1998).}
\label{fig:nuspectrum}
\end{figure}

What can we learn from this bit of history?  When Ray and I wrote our
PRL papers arguing that a chlorine detector of 600 tons could observe
solar neutrinos, we never discussed the possibility of using neutrinos to
learn about particle physics. The only motivation we gave was ``...to
see into the interior of a star and thus verify directly the
hypothesis of nuclear energy generation in stars.'' [PRL 12, 300 (1964)].  

Why did we not discuss using neutrinos for particle physics?  Frankly,
because we never thought about it. And even if we had, we would have
known better than to mention it to our particle physics friends. Bruno
had the insight and the vision and indeed the courage to argue that
astronomical neutrinos could potentially give us unique information
about neutrino characteristics.  His paper is all the more
remarkable because it was published a year before the first results of
the chlorine experiment showed that the rate Ray observed was less
than our calculated rate.

We learn from these events that pioneering experiments can lead to
important results in areas that are unanticipated.  We will come back
to this conclusion at the end of this talk.

\subsection{Standard Model Predictions}
\label{subsec:standardmodel}

Figure~\ref{fig:nuspectrum} shows the calculated solar neutrino
spectrum predicted by the Standard solar model.  The percentage errors
are the calculated $1\sigma$ uncertainties in the predicted fluxes,
based upon the published errors of the measured quantities and on many
calculations of standard solar models.  As you will hear from the
talks in the later parts of this morning session, the total
intensities and the energy spectra shown in Fig.~\ref{fig:nuspectrum}
are now widely used to interpret, and indeed to plan, solar neutrino
experiments such as those discussed in today's sessions: chlorine,
Super-Kamiokande, SNO, SAGE, GALLEX, GNO, and BOREXINO.

\begin{figure}[!htb]
\hglue-.48in\psfig{figure=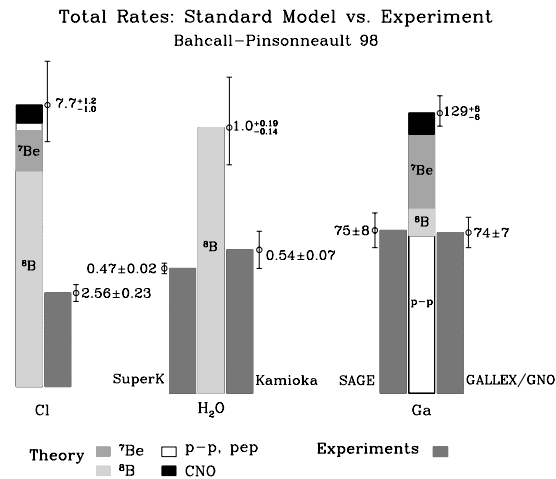,width=3.7in,angle=270}
\vglue-.5in
\caption[]{Comparison of measured rates and standard-model (BP98) predictions
for six solar neutrino experiments.  The unit for the radiochemical
experiments (chlorine and gallium) is SNU ($10^{-36}$ interactions per
target atom per sec); the unit for the
water-Cerenkov experiments (Kamiokande and Super-Kamiokande) is the
rate predicted by the standard solar model plus standard electroweak
theory. The experimental results are described by Lande, Suzuki,
Gavrin, and Belotti in these proceedings.
\label{fig:theoryvsexp}}
\end{figure}

Figure~\ref{fig:theoryvsexp} compares the calculated versus the
measured rates for the six solar neutrino experiments for which
results have been reported.  Assuming nothing happens to the neutrinos
after they are created, the measured rates range from $33$\% $\pm 5$\%
of the calculated rate (for chlorine) to $58$\% $\pm 7$\%.  As is now
well known, the observed rates cannot be fit (at a C.L. of about 99\%)
with any linear combination of undistorted solar neutrino energy spectra.

Today we know that there are three reasons that the calculations of
solar neutrino fluxes are robust: 1) the availability of precision
measurements and precision calculations of input data that have been
gradually refined over four decades; 2) the intimate connection
between neutrino fluxes and the measured solar luminosity; and 3) the
measurement of the helioseismological frequencies of the solar
pressure-mode ($p$-mode) eigenfrequencies.

Could the solar model calculations be wrong by enough to explain the
discrepancies between predictions and measurements shown in
Fig.~\ref{fig:theoryvsexp}? 
Helioseismology, which confirms predictions of the
standard solar model to high precision, suggests that the answer is
 ``No.''

\begin{figure}[!htb]
\centerline{\psfig{figure=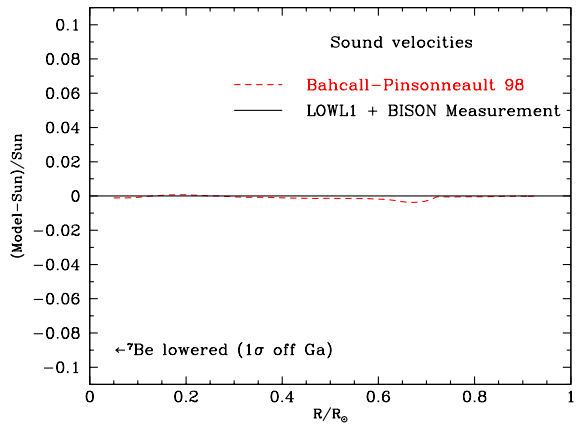,width=3.3in,angle=270}}
\vglue-.4in
\caption[]{Predicted versus Measured Sound Speeds.  This figure shows
the excellent agreement between the calculated (solar model BP98,
Model) and the measured (Sun) sound speeds, a fractional difference of
$0.001$ rms for all speeds measured between $0.05 R_\odot$ and $0.95
R_\odot$. The vertical scale is chosen so as to emphasize that the
fractional error is much smaller than generic changes in the model,
$0.09$, that might significantly affect the solar neutrino
predictions.  The measured sound speeds are from S. Basu et al.,
Mon. Not. R. Astron. Soc. 292, 234 (1997); The figure is taken from
BP98. \protect\label{fig:diffmodelbest}}
\end{figure}

Figure~\ref{fig:diffmodelbest} shows the fractional differences between the
most accurate available sound speeds measured by
helioseismology and sound speeds calculated with our
best solar model (with no free parameters).  The horizontal line
corresponds to the hypothetical case in which the model predictions
exactly match the observed values.  The rms fractional difference
between the calculated and the measured sound speeds is
$1.1\times10^{-3}$ for the entire region over which the sound speeds
are measured, $0.05R_\odot < R < 0.95 R_\odot$.  In the solar core,
$0.05R_\odot < R < 0.25 R_\odot$ (in which about $95$\% of the solar
energy and neutrino flux is produced in a standard model), the rms
fractional difference between measured and calculated sound speeds is
$0.7\times10^{-3}$.

The arrow in Fig.~\ref{fig:diffmodelbest} shows how different the
solar model sound speeds would have to be from the observed sound
speeds if one wanted to use solar physics to reduce the $^7$Be
neutrino flux. The position of the arrow is fixed by artificially
reducing the predicted $^7$Be neutrino flux that is not observed in
the gallium experiments, SAGE and GALLEX plus GNO (see
Fig.~\ref{fig:theoryvsexp}). if the $p-p$ neutrinos are present.
Remember, we believe we can calculate the $p-p$ flux to $\pm 1$\%.

The discrepancy with the hypothetical new solar physics was
estimated by using the temperature dependence of the $^7$Be neutrino
flux ($\propto T^{10}$) and the sound speeds ($\propto T^{1/2}$).  The
agreement with the hypothetical solar physics is more than $100$ times
worse than the agreement with the Standard Model physics.

Figure~\ref{fig:diffmodelbest} has contributed 
to the consensus view that the experimental results shown in
Fig.~\ref{fig:theoryvsexp} require new particle physics for their
explanation.

\subsection{Summing up and looking ahead}
\label{subsec:summingup}

I want now to look back and then  
look ahead. I will begin by giving my view
of the principal accomplishments of solar neutrino research to date
(Sect.~\ref{subsubsec:achievements}).  Then I will discuss two of
the expected highlights of the next decade of solar neutrino research,
the measurement of the neutral current to charge current ratio for
$^8$B neutrinos ((Sect.~\ref{subsubsec:doubleratio}) and the
detection of solar neutrinos with energies less than 1 MeV
((Sect.~\ref{subsubsec:below1mev}). 

\subsubsection{Principal achievements}
\label{subsubsec:achievements}

What are the principal achievements of the first four decades of solar
neutrino research?  I give below my personal list of the `top three
achievements.' 

$\bullet$ {\bf Solar neutrinos have been detected.}  The chlorine,
Kamiokande, Super-Kamiokande, GALLEX, SAGE, GNO, and SNO experiments
have all measured solar neutrino events. This is the most important
achievement. The detection of solar neutrinos shows empirically that
the sun shines by the fusion of light elements.

$\bullet$ {\bf Evidence for new physics has been found.} For more than
thirty years, beginning with the fact that Ray's first measurements in
1968 
indicated a flux lower than the standard model predictions, we have
had evidence for new physics in the solar neutrino arena. This
evidence has steadily deepened as new solar neutrino experiments have
confirmed and extended the neutrino discrepancies and helioseismology
has confirmed the standard solar model.  The fact that neutrino
oscillations have now been observed in atmospheric neutrino phenomena
further strengthens the case that oscillations occur for solar
neutrinos. We are still looking for a `smoking gun' single effect that
shows up in just one solar neutrino experiment, rather than combining
the results of two or more different experiments. I will discuss some
possibilities below.

$\bullet$ {\bf Neutrino fluxes and energy spectra are approximately as
predicted by the standard solar model.}  If you had told me in 1964
that six solar neutrino experiments would give results within a factor
of three of the predicted standard model results, I would have been
astonished and delighted. This is especially so considering that the
crucial $^8$B neutrino flux depends upon the 25th power of the central
temperature of the sun. This agreement exists without making any
corrections for neutrino oscillations.

 If we correct the observed solar neutrino event rates for the effects
of neutrino oscillations using the six currently allowed two-neutrino
oscillation scenarios, the inferred $^8$B neutrino flux at the source
is rather close to the best-estimate predicted flux. At the 99\% CL,
one infers (see hep-ph/9911248):

\begin{equation}
0.55 \leq \phi(^8B)/{(\rm Standard ~ prediction)} \leq 1.32, 
\label{eqn:b8limits}
\end{equation}
which is a
slightly tighter range than the $3\sigma$ prediction of the standard
solar model.

\subsubsection{SNO and the [NC]/[CC] ratio}
\label{subsubsec:doubleratio}

\begin{figure}[!h]
\centerline{\psfig{figure=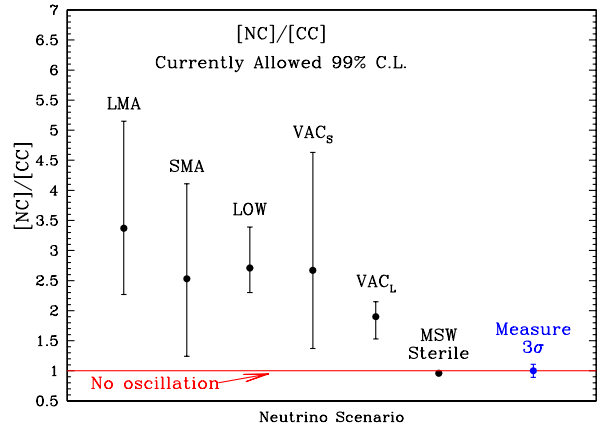,width=3.3in,angle=270}}
\caption[]{\small The neutral current to charged current double ratio,
[NC]/[CC] .  The standard model value for [NC]/[CC] is $1.0$.  The
figure shows, for a 5 MeV threshold for the CC measurement, the
predicted double ratio of Neutral Current to Charged Current for
different neutrino scenarios.  The solid error bars represent the
$99$\% C.L. for the allowed regions of the six currently favored
neutrino oscillation solutions.  The dashed error bar labeled
``Measure $3\sigma$'' represents the net estimated uncertainty in
interpreting the measurements, including the energy resolution,
energy scale, $^8$B neutrino energy spectrum, neutrino cross section,
counting statistics, and the $hep$ flux. This is Fig. 7a of Bahcall,
Krastev, and Smirnov, hep-ph/0002293.
\label{fig:double}}
\end{figure}

Figure~\ref{fig:double} shows the predictions of the currently allowed
neutrino oscillation solutions for the double ratio, [NC]/[CC], of
neutral current to charged current event rates in the deuterium
detector SNO.  Art McDonald will describe later this morning the
experimental characteristics of this great observatory and outline for
us the extensive program of SNO measurements. The important message of
Fig.~\ref{fig:double} is that all of the currently allowed oscillation
solutions for active neutrinos predict a value for the double ratio
that is different from the no oscillation value of 1.0 by at least
nine times the estimated non-statistical measurement uncertainty.  

We all eagerly look
forward to this crucial and decisive measurement.

\subsubsection{Solar neutrinos below 1 MeV}
\label{subsubsec:below1mev}

More than $98$\% of the calculated standard model solar neutrino flux
lies below $1$ MeV. The rare $^8$B neutrino flux is the only solar
neutrino source for which measurements of the energy have been made,
but $^8$B neutrinos constitute a fraction of less than $10^{-4}$ of
the total solar neutrino flux.

The great challenge of solar neutrino astronomy is to measure neutrino
fluxes below $1$ MeV. We must develop experiments
that will measure the $^7$Be neutrinos (energy of $0.86$ MeV) and the
fundamental $p$-$p$ neutrinos ($< 0.43$ MeV).  A number of promising
possibilities were discussed at the LowNu workshop that preceded this
conference. The BOREXINO observatory, which can detect $\nu-e$
scattering, is the only approved solar neutrino experiment which can
measure energies less than $1$ MeV.

The $p$-$p$  neutrinos are overwhelmingly the most abundant source of
solar neutrinos, carrying about $91$\% of the total flux according to
the standard solar model. The $^7$Be neutrinos constitute about $7$\%
of the total standard model flux.

We want to test and to understand neutrino oscillations with high
precision using solar neutrino sources. To do so, we have to measure the
neutrino-electron scattering rate with $^7$Be neutrinos, as will be
done with the BOREXINO experiment, and also the CC
(neutrino-absorption) rate with $^7$Be neutrinos (no approved
experiment). With a neutrino line as provided by $^7$Be
electron-capture in the sun, unique and unambiguous tests of neutrino
oscillation models can be carried out if one measures both the
charged-current and the neutral current reaction rates.

I believe that we have calculated the flux of $p$-$p$ neutrinos produced in
the sun to an accuracy of $\pm 1$\%. This belief should be tested
experimentally. Unfortunately, we do not yet have a direct measurement
of this flux. The gallium experiments only tell us the rate of capture
of all neutrinos with energies above $0.23$ MeV.

The most urgent need for solar neutrino research is to develop a
practical experiment to measure directly the $p$-$p$ neutrino flux and
the energy spectrum of electrons produced by  weak interactions with
$p$-$p$ neutrinos.  Such an experiment can be used to test the precise
and fundamental standard solar model prediction of the \hbox{$p$-$p$} neutrino flux.
Moreover, the currently favored neutrino oscillation solutions all
predict a strong influence of oscillations on the low-energy flux of
$\nu_e$.

\begin{figure}[!htb]
\centerline{\psfig{figure=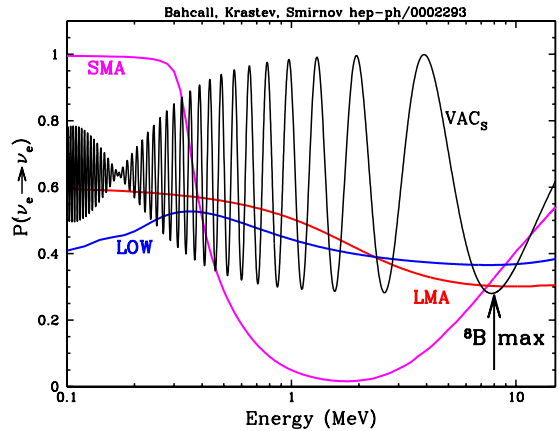,width=3.6in,angle=270}}
\vglue-.6in
\caption[]{Survival probabilities for MSW solutions.
The figure presents the yearly-averaged 
survival probabilities for an electron neutrino that is  created
in the sun to remain an electron neutrino 
 upon arrival at the Super-Kamiokande
detector. 
\label{fig:survival}}
\end{figure}

Figure~\ref{fig:survival} shows the calculated neutrino survival
probability as a function of energy for three global best-fit MSW
oscillation solutions.  You can see directly from this figure why we
 need accurate measurements for the $p$-$p$ and $^7$Be neutrinos.
The currently favored solutions exhibit their most characteristic and
strongly energy dependent features below $1$ MeV.  Naturally, all of
the solutions give similar predictions in the energy region, $\sim 7$
MeV, where the Kamiokande and Super-Kamiokande data are best.  
 The survival probability shows a strong change with
energy below $1$ MeV for all the solutions, whereas in the region above $5$ MeV (accessible
to Super-Kamiokande and to SNO) the energy dependence of the survival
probability is at best modest.
 
The $p$-$p$ neutrinos are the gold ring of solar neutrino astronomy.
Their measurement will constitute a simultaneous and critical test of
stellar evolution theory and of neutrino oscillation solutions.

\section{Extragalactic neutrinos}
\label{sec:extragalactic}

\begin{figure}[!thb]
\centerline{\psfig{figure=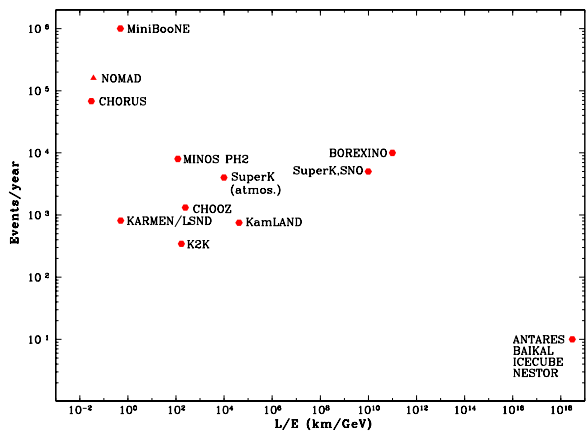,width=3.3in,angle=270}}
\caption[]{\small Very longbaseline neutrino oscillation
experiments. The figure shows that experiments such as ANTARES,
BAIKAL, ICECUBE, and NESTOR, which may detect high-energy neutrinos
from distant gamma-ray bursts, have extraordinary sensitivity to
vacuum neutrino oscillations.  Neutrinos of $10^{5}$ GeV from
gamma-ray bursts located at cosmological distances were used to locate
the positions of ANTARES, BAIKAL, ICECUBE, and NESTOR in the figure.
\label{fig:lovere}}
\end{figure}

Experimentalists often like to describe the power of their experiments
in terms of the expected or observed number of events per year and $L/E$, where
$L$ is the distance between the accelerator and the detector and $E$
is the beam energy. The quantity $L/E$ determines, together with the
square of the mass difference, the survival probability for vacuum
neutrino oscillations. More generally, $L/E$ represents the time of
flight in the rest frame of the particle, the time for rare events to
occur.

Figure~\ref{fig:lovere} shows the extraordinary sensitivity to neutrino
oscillation of experiments like ANTARES, BAIKAL, ICECUBE, and NESTOR that can
detect neutrinos from distant extragalactic sources. The accelerator
experiments that will be discussed at Neutrino 2000 lie in the
left-hand side of Fig.~\ref{fig:lovere}, $L/E < 10^{4}$ km/GeV .  Solar
neutrino experiments like Super-Kamiokande, SNO, and BOREXINO can reach
to $10^{10}$ km/GeV and, for the lower energy experiments, even
$10^{11}$ km/GeV.  Extragalactic sources such as gamma-ray-bursts
(GRBs) have such a long baseline ($\sim 10^{10}$ lyrs) that the new
generation of extragalactic experiments,  ANTARES, BAIKAL, ICECUBE, and
NESTOR will extend to the right-hand side of Fig.~\ref{fig:lovere}, to
$L/E ~>~10^{18}$~km/GeV. 

I want to say a few words about the possibilities for detecting GRB
neutrinos, which will be discussed in more detail in these proceedings
by Eli Waxman.  I believe that GRBs offer the best chance for
detecting extragalactic neutrinos among all the known sources of
astronomical photons.

The phenomenology of the photons observed from
gamma-ray bursts is now relatively well understood. Many different
types of observations have been carried out and the results are well
summarized by the expanding fireball model. Using this model, one can
work out the flux of neutrinos from shocks.  

\begin{figure}[!t]
\centerline{\psfig{figure=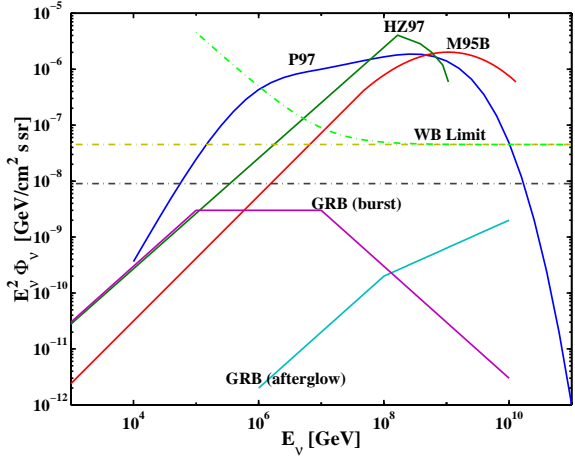,width=3in}}
\caption[]{The Waxman-Bahcall upper bound on muon neutrino intensities
($\nu_\mu + \bar\nu_\mu$). This figure is from Bahcall and Waxman, hep-ph/9902383.  The numerical value of the bound assumes
that $100$\% of the energy of protons is lost to $\pi^+$ and $\pi^0$
and that the $\pi^+$ all decay to muons that also produce neutrinos.
The dot--dash line gives the upper bound corrected for neutrino energy
loss due to redshift and for the maximum known  evolution (QSO
or star-formation evolution).  The lower line is obtained assuming no
evolution. The solid curves show the predictions of representative AGN
jet models taken from the earlier papers of Mannheim (marked M95B in
the figure), Protheroe~(P97), and Halzen and Zas~(HZ97).  The AGN
models were normalized so that the calculated gamma-ray flux from
$\pi^0$ decay fits the observed gamma-ray background. }
\label{fig:jetbound}
\end{figure}

Figure~\ref{fig:jetbound} shows the neutrino energy spectra that
Waxman and I have estimated to be produced by GRBs, both from the
direct burst (energies $\sim 10^{6}$ GeV) and from the afterglow
(energies $\sim 10^{8}$ GeV to $\sim 10^{19}$ GeV).  The  observed
population of GRBs should give rise to $\sim 10$ events per ${\rm
km^2}$ per year from neutrinos with characteristic energies of order
$10^{14}$ eV. We shall hear on the last day of this
conference that the calculated GRB flux may be detectable in ANTARES, ICECUBE, or NESTOR.
 The fundamental assumption used in calculating the GRB neutrino flux 
is that GRBs produce the observed flux of high-energy cosmic rays, an
assumption for which Eli Waxman has provided a strong plausibility
argument.

 GRBs occur at modest to large redshifts.  We know the time of the
explosion to an accuracy $\sim 10$ sec (from the gamma
rays). Therefore, GRBs can be used to test special relativity to an
accuracy of $1$ part in $10^{16}$ and to test the weak equivalence
principle to an accuracy of $1$ part in $10^6$.  If special relativity
is right, the photons and the neutrinos should arrive at the same time
(to an accuracy of about $10$ sec, the duration of the burst). If the
weak equivalence principle is valid, the arrival times of neutrinos
(which traverse significant gravitational potentials) from distant
sources should be independent of neutrino flavor.

GRBs can also be used to probe the weak interactions to an
extraordinary level of precision. Gamma-ray bursts 
are expected to produce only $\nu_e$ and
$\nu_{\mu}$. The large area detectors of extragalactic neutrinos are
in principle sensitive to vacuum neutrino oscillations with mass
differences as small as $\Delta m^2 ~\geq~ 10^{-17}~ {\rm
eV^2}$~~(from $\nu_\mu \rightarrow \nu_\tau$).

Not everything is encouraging in Fig.~\ref{fig:jetbound}.  The figure
also shows the upper limit that is allowed for astrophysical neutrino
production from $(\gamma,\pi)$ interactions on high energy
protons. The upper limit is established by using the observed cosmic
ray flux of high energy protons.  Prior to the recognition of this
limit a number of authors had suggested much more optimistic models
(also shown in the figure), that were normalized by fitting
$\pi^0$ decay to the observed gamma-ray background. 

\section{Goals for Astrophysical neutrinos: 2000-2010}
\label{sec:goals}

It seems to me that we have three principal goals for this next decade.
\begin{quotation}
\noindent\raggedright
\begin{itemize}
\item Determine the mixing angles and mass differences that
are important for solar neutrino phenomena.

\item Test precisely stellar evolution by observing $p-p$ and
$^7$Be neutrinos, and by determining the total flux of $^8$B neutrinos.

\item Discover extragalactic neutrinos, perhaps from
gamma-ray bursts.
\end{itemize}
\end{quotation}

From time to time, friends ask me to compare the search for solar
neutrinos with the search for neutrinos from GRBs. They are very
different. From photon studies, we know more observationally
about the sun than about any other astronomical source, certainly much
more than about the mysterious GRBs. Moreover the sun is in the
simplest stage of stellar evolution, in quasi-static equilibrium with
a characteristic time scale for evolution of $10^9$ yr ($10^{16}$ s ). We
do not even know the energy source of GRBs. We do know that GRBs are
far from equilibrium, evolving explosively on a time scale of order
 $10^{-3}$ s.

We want to do extragalactic neutrino astronomy because it is truly an
exploration of the universe. We do solar neutrino astronomy to test
fundamental theories of physics and astronomy. But, perhaps solar
neutrino research and extragalactic neutrino research may in the end
share a fundamental characteristic: surprise. Remember, that we
undertook solar neutrino research to test stellar evolution and
unexpectedly (at least for everybody except Bruno Pontecorvo) we
found evidence for new neutrino physics.

In a sense, we are returning to our original goal in neutrino
astronomy, but by a round-about path. We must first understand
neutrino oscillation phenomena in order to be able to use solar
neutrino observations to test precisely the theory of stellar
evolution, our original goal. Perhaps with extragalactic astronomy we
will participate in a similar cycle of  astronomical exploration and
physical clarification.  T. S. Elliot in `The Four Quartets' described
the cycle succinctly and beautifully:

\begin{quote}

We shall not cease from exploration

And the end of all of our exploring

Will be to arrive where we started 

And know the place for the first time.

\end{quote}

\section*{Acknowledgments}

I acknowledge support from NSF grant \#PHY95-13835.

\end{document}